\documentclass{appolb}
\usepackage{epsfig}
\usepackage[dvips]{color}
\begin{document}
\large

\title{Mirror objects in the solar system?}

\author{Z.~K.~Silagadze
%\vspace*{3mm} \\
\address{
Budker Institute of Nuclear Physics, 630 090 Novosibirsk, Russia}
}
%\date{}

\maketitle

\begin{abstract}
This talk was given at the Tunguska-2001 international conference but it is
not about the Tunguska event. Instead we tried to give some flavor of
mirror matter, which is predicted to exist if parity is an unbroken symmetry 
of nature, to non-experts. The possible connection of the mirror matter
ideas to the Tunguska phenomenon was indicated by Foot and Gninenko some 
time ago and was elaborated by Foot in the separate talk at this conference.
If the mirror world interpretation of the Tunguska like events is indeed
correct then the most fascinating (but very speculative) possibility is that
some well known celestial bodies with strange properties are in fact made 
mostly from mirror matter, and so maybe the mirror world was discovered 
long ago and we just have not suspected this!    
\end{abstract}
\PACS{11.30.Er, 95.30.-k, 95.35.+d}

\section*{}
This conference is devoted to the 1908 Tunguska mysterious event. Human 
beings like mystery stories very much. Maybe the greatest mystery being
our very ability to be so curious. It seems that the strange aspiration
for unraveling mysteries and even stranger belief that the truth really
exists for every case is hardwired in our brains. This passion for knowledge
is powerful enough to compete with other human passions and makes possible
the existence of substantial science despite the fact that scientists, just
like other human beings, are subject to various human weaknesses not
compatible with genuine science like arrogance, vanity and unfairness.

It is not easy to find the truth about the event so old and so enigmatic.
Thus it is not surprising that numerous hypotheses were suggested for
explanation to what happened in the Central Siberia near the Podkamennaya  
Tunguska River in the early morning hours of June 30, 1908 \cite{1}. A new 
hypothesis is considered in Foot's talk at this conference \cite{2} 
according to which Tungus tribesmen and Russian fur traders had witnessed
an atmospheric explosion of some mirror meteoroid. What is mirror meteoroid?
It is a meteoroid made from the mirror matter. And what follows is an attempt
to explain to you the meaning of words ``mirror matter''. While describing
certainly exotic things, we will try to follow the advice ``Be open-minded, 
but not so open-minded that your brains fall out.''  

The main motivation for the mirror-world comes from the symmetry argument:
the existence of mirror matter makes the world left-right symmetric.
What is left and what is right at the fundamental, that is quark and lepton
level, needs some explanation. You know that many physical quantities are 
vectors, like velocity or acceleration. The main characteristic property of 
vectors is their transformation law under rotations. For example, if one 
rotates a radius-vector $\vec{r}=(x,y,z)$ by the angle $\theta$ around the
$z$-axis its $x$- and $y$-coordinates will change according to
\begin{eqnarray} 
x^\prime&=&\;\;\;\,\cos{\theta}~x+\sin{\theta}~y  \nonumber \\ 
y^\prime&=&-\sin{\theta}~x+\cos{\theta}~y
\label{eq1} \end{eqnarray}
Any other vector also transforms like this. Now the transformation law above
shows that a vector remains unchanged under $360^\circ$-rotation. Therefore
vectors can not be the most fundamental objects because the 
$360^\circ$-rotation is not an identity transformation and the most 
fundamental objects are expected to change under such rotations. The last 
assertion does look strange, does it not? Why a $360^\circ$-rotation is not
an identity transformation? In fact it is, but only for isolated objects.
In general something changes in this world when somebody makes a full turn
on his heels. Our ancestors intuitively always understood this. In fairy
stories one can find quite often an assertion like this: ``The magician 
turned around on his heels and turned into a mouse.'' If you do not believe 
in fairy stories maybe the following demonstration by Dirac \cite{3} will be
more convincing for you. Take a triangle made from some hard material (Dirac
himself used a pair of scissors) and attach elastic strings to its vertexes.
Fix other ends of the strings, for example as shown in the figure below
\begin{figure}[htb] 
   \begin{center} \mbox{\epsfig{figure=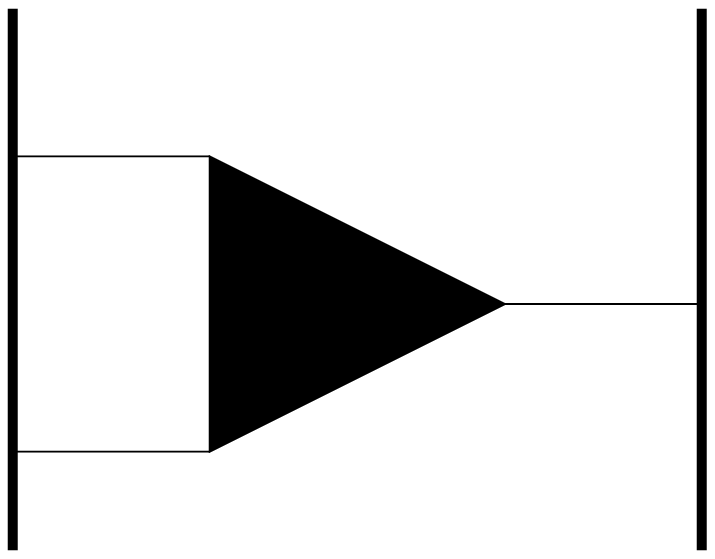 , height=3.0cm}} 
     \end{center}   
\end{figure}

\noindent
Use the sole string as the rotation axis and turn the triangle around it
by one full turn. The other two strings will become twisted. Now keep the 
triangle fixed and try to remove the twist by manipulating the elastic 
strings. You failed to achieve this? Not surprising because it can be
proved \cite{4} that it is impossible. But more surprises are waiting to you.
Go ahead and turn again the triangle by one full turn in the same direction.
The strings will become twice more twisted as the triangle has made two full 
turns. However this time it is possible to untwist strings by taking them
over and around the fixed triangle. This fact demonstrates clearly that 
only the $720^\circ$-rotation and not the $360^\circ$-rotation is the 
identity transformation. But to really believe this you should perform the
above described exercise by yourself.

Actually there is another way to demonstrate the distinction between  
$360^\circ$- and $720^\circ$-rotations. For this demonstration you do not 
need to prepare any equipment at all except a cup of coffee. It turns out
that our arms are properly designed for the trick which demonstrates
the distinction between $360^\circ$- and $720^\circ$-rotations. The trick
was invented by Balinese candle-dancers long ago and performed countless
times without realizing any deep mathematics behind it. In presence of
the more scientifically trained audience the trick was firstly performed
by Feynman during his 1986 Dirac memorial lecture \cite{5}. Now you can
try this Balinese candle-dance trick by yourself using the following 
instructions given by Burton \cite{6}: ``You hold the coffee cup with your 
right hand underneath it, straight out in front of you. Now bring it left, 
under your underarm, awkwardly around front with your elbow straight up in 
the air. That's 360 degrees, and you're a pretzel. Keep going around 
counterclockwise, this time swinging your arm around over your head. At 720 
degrees the coffee cup is back where it started, unspilled, and your arm is 
straight once more. Keep going round and round until you believe it.''

The trick even has a technical application. According to Hansen \cite{3},
in 1971 D.~A.~Adams patented in USA a solution to the problem of transferring
electrical current to a rotating plate without the wires being entangled 
based on the Balinese candle-dance trick. For another interesting application
of a Balinese candle dance,  at 7200 revolutions per minute, in a medical 
centrifuge see Burton's story \cite{6}. Let us also indicate some other 
literature sources \cite{7} where you can find further discussion of 
$360^\circ$-rotations.

To find more fundamental objects behind vectors let us note that from a 
vector one can make $2\times2$ traceless matrix by using Pauli matrices.
For example, the radius-vector is associated with the matrix
$$X=x\sigma_1+y\sigma_2+z\sigma_3=\left ( \begin{array}{cc} z & x-iy \\
x+iy & -z \end{array} \right ). $$ 
Under rotations (\ref{eq1}) the matrix $X$ transforms like this
\begin{equation}
X^\prime =UXU^+,
\label{eq2}\end{equation}
where
\begin{equation}
U=e^{i\sigma_3\frac{\theta}{2}}=\left ( \begin{array}{cc} e^{\frac{i\theta}
{2}} & 0 \vspace*{2mm} \\ 0 & e^{-\frac{i\theta}{2}} \end{array} \right ). 
\label{eq3}\end{equation}
Note that the transformation matrix $U$ depends now on $\frac{\theta}{2}$ 
and thus changes sign under $360^\circ$-rotation! Let us decompose the 
matrix $X$ into simpler building blocks:
\begin{equation}
X=\xi\xi^TC=\left ( \begin{array}{c} \xi_1 \\ \xi_2  \end{array} \right )
\left ( \begin{array}{cc} \xi_1 & \xi_2  \end{array} \right )
\left ( \begin{array}{cc} 0 & 1 \\ -1 & 0 \end{array} \right )=
\left ( \begin{array}{cc} -\xi_1 \xi_2 & \xi_1^2 \\ -\xi_2^2 & 
\xi_1 \xi_2 \\ \end{array} \right ) ,
\label{eq4}\end{equation}
where $\xi$ is an object with 2 complex components 
$$\xi=\left ( \begin{array}{c} \xi_1 \\ \xi_2  \end{array} \right )$$
and the presence of the ``charge conjugation matrix'' $C=i\sigma_2$ is 
necessary to make the r.h.s. of (\ref{eq4}) traceless (by this condition
the constant matrix $C$ is determined uniquely up to normalization).
Because $C\sigma_3^+=-\sigma_3^TC$ we will have $CU^+=U^TC$ and 
$$X^\prime=U\xi\xi^TCU^+=U\xi\xi^TU^TC=(U\xi)(U\xi)^TC.$$
Therefore the transformation law for the object $\xi$ is
\begin{equation}
\xi^\prime=U\xi
\label{eq5}\end{equation}
and this object changes sign under $360^\circ$-rotations. Such objects, some
kind of square root of vectors, are called spinors. And quarks and leptons 
are spinors. One can think that this sign change under $360^\circ$-rotations 
should be irrelevant. Especially if you remember that wave function phase does
not matter in quantum mechanics. Indeed a $360^\circ$-rotation does not matter
for {\it isolated} spinor. But if this spinor is a part of a bigger system 
this minus sign matters a lot: it leads to the Pauli exclusive principle 
\cite{5,6}. Hence all the chemistry and our very existence are based on this 
minus sign! 

Now it is time to ask how the spinors transform under Lorentz transformations.
Equation (\ref{eq5}) suggests the following general transformation law for 
spinors
$$\xi^\prime=\exp{\left \{i\sum\limits_{i=3}^3({\bf J_i}\theta_i+
{\bf K_i}\varphi_i)
\right\}}\xi,$$
where ${\bf J_i}=\frac{1}{2}\sigma_i$ are generators of spatial rotations and
${\bf K_i}$ are Lorentz boost generators. The boost is characterized by the 
parameters $\varphi_i$. For example, for the Lorentz boost along 
$x$-direction one has
\begin{eqnarray} 
x^{0\prime}&=&\cosh{\varphi}~x^0+\sinh{\varphi}~x  \nonumber \\ 
x^\prime&=&\sinh{\varphi}~x^0+\cosh{\varphi}~x
\label{eq6} \end{eqnarray}
where $\cosh{\varphi}=\gamma$ determines the $\gamma$-factor of the boost.
Formulas like (\ref{eq1}) and (\ref{eq6}) in fact determine explicit forms
of ${\bf J_i}$ and ${\bf K_i}$ generators for 4-vectors and hence their 
commutators. It turns out \cite{8} that the commutation relations can be 
written in the form
\begin{equation}
[{\bf A_i^{\pm}},{\bf A_j^{\pm}}]=i\epsilon_{ijk}{\bf A_k^{\pm}}
,\;\; [{\bf A_i^+},{\bf A_j^-}]=0,
\label{eq7} \end{equation}
where ${\bf A_i^{\pm}}=\frac{1}{2}({\bf J_i}\pm i{\bf K_i})$. 
These commutation relations show that the Lorentz group is locally identical
to the $SU(2)\times SU(2)$ group. Therefore its representations are labeled
by two angular momenta $j_-$ and $j_+$. All such $(j_-,j_+)$ representations
can be constructed from the two fundamental spinor representations 
$(\frac{1}{2},0)$ and $(0,\frac{1}{2})$. Therefore we have two kinds of 
spinors. The left spinor $(\frac{1}{2},0)$ transforms non-trivially under 
the left $SU(2)$ formed by the ${\bf A_i^-}$ generators while remaining 
unchanged under the right $SU(2)$ formed by the ${\bf A_i^+}$ generators. 
For the right spinor  $(0,\frac{1}{2})$ roles of the left and right 
$SU(2)$ factors of the Lorentz group are interchanged. Left spinor is 
annihilated by the ${\bf A_i^+}=\frac{1}{2}({\bf J_i}+i{\bf K_i})$ generators.
Therefore for such spinor ${\bf K_i}=i{\bf J_i}=\frac{i}{2}\sigma_i$, 
because ${\bf J_i}=\frac{1}{2}\sigma_i$ for spinor 
representation. Analogously for right spinor ${\bf K_i}=-\frac{i}{2}\sigma_i$.
Thus under Lorentz boosts left spinor $\xi_L$ and right spinor $\xi_R$
transform as 
\begin{equation}
\xi_L^\prime =e^{-\frac{1}{2}\vec{\sigma}\cdot\vec{\varphi}}\xi_L, \;\;\;
\xi_R^\prime =e^{\frac{1}{2}\vec{\sigma}\cdot\vec{\varphi}}\xi_R. 
\label{eq8} \end{equation}
It turns out \cite{8} that for massless spinors the projection of their spin
on their momentum direction is negative for left spinors and positive for
right spinors. So we can think about left and right spinors as some analogs
of left-handed and right-handed screws.

Let ${\bf P}$ be space inversion (or parity) operator
\begin{equation}
{\bf P}:~(x_0,x,y,z)\rightarrow (x_0,-x,-y,-z).
\label{eq9} \end{equation}
How do spinors transform under parity? Inspecting 4-vector transformation 
laws under Lorentz boosts (\ref{eq6}) and under parity (\ref{eq9}) one can
find that ${\bf P}$ anticommutes with the boost generators ${\bf K_i}$. 
But then it is impossible to realize the parity operator by a $2\times 2$
matrix in the $\xi_L$ or $\xi_R$ space because no $2\times 2$ matrix 
anticommutes with all Pauli matrices. The analogy with screws hints a rescue.
Under space inversion left-handed screw goes into right-handed screw and 
vice versa. Therefore the parity operator should transform left spinors into
right spinors and 
vice versa. Hence to have a spinor realization of the Lorentz group extended
by parity one should unify $\xi_L$ and $\xi_R$ spinors into a one 4-component
object (Dirac spinor)
$$\psi=\left( \begin{array}{c} \xi_R \\ \xi_L \end{array} \right).$$
Then in the space spanned by $\psi$ spinors there is enough room to realize 
both  Lorentz boost generators ${\bf K_i}$ and the parity operator ${\bf P}$:
$${\bf K_i}=\left (\begin{array}{cc} -\frac{i}{2}\sigma_i & 0 \\ 0 &
\frac{i}{2}\sigma_i \end{array} \right),\;\;\; {\bf P}=\left 
(\begin{array}{cc} 0 & 1 \\ 1 & 0 \end{array} \right),\;\;\; 
{\bf K_i}{\bf P}=-{\bf P}{\bf K_i}\,. $$
A bit more information about the fundamental nature of left and right spinors
and we will be ready to discuss their connection to the mirror matter idea.
Using
$$\cosh{\frac{\varphi}{2}}=\sqrt{\frac{1+\gamma}{2}}, \;\;\;
\sinh{\frac{\varphi}{2}}=\sqrt{\frac{\gamma-1}{2}}, \;\;\;
\gamma=\frac{E}{m} $$
we get
$$e^{\frac{1}{2}\vec{\sigma}\cdot\vec{\varphi}}=\cosh{\frac{\varphi}{2}}+
\frac{\vec{\sigma}\cdot \vec{p}}{|\vec{p}|}\sinh{\frac{\varphi}{2}}=
\frac{E+m+\vec{\sigma}\cdot \vec{p}}{\sqrt{2m(E+m)}}.$$
Therefore equations (\ref{eq8}) indicate
\begin{equation}
\xi_R(p)=\frac{E+m+\vec{\sigma}\cdot\vec{p}}
{\sqrt{2m(E+m)}}~\xi_R(0), \;\;\; \xi_L(p)=\frac{E+m-\vec{\sigma}\cdot\vec{p}}
{\sqrt{2m(E+m)}}~\xi_L(0),
\label{eq10} \end{equation}
where $\xi_R(0)$ and $\xi_L(0)$ are rest-frame spinors. But you can not tell
whether a screw is left-handed or right-handed if the direction the screw
points is unknown. So for the rest frame spinors it should be impossible to
distinguish left spinors from right spinors and one should have \cite{8,9}
\begin{equation}
\xi_R(0)=\xi_L(0).
\label{eq11} \end{equation}
Taking this into account allows to rewrite (\ref{eq10}) as
$$(E-\vec{\sigma}\cdot\vec{p})~
\xi_R(p)=m\xi_L(p),\;\;\; (E+\vec{\sigma}\cdot\vec{p})~\xi_L(p)=m\xi_R(p).$$
But this is nothing but the Dirac equation for the Dirac spinor $\psi$
$$(\hat p-m)\psi=0,$$
where the Dirac matrices are given in the chiral representation
$$\gamma_0=\left ( \begin{array}{cc} 0 & 1 \\ 1 & 0 \end{array} \right ),
\;\;\; \gamma_i=\left ( \begin{array}{cc} 0 & -\sigma_i \\
\sigma_i & 0 \end{array} \right ). $$
You surely know about the central role the Dirac equation plays in our 
understanding of Nature. And we have just discovered that the Dirac equation
is merely assertion that for the spinor particle at rest one can not tell
whether it is left-handed or right-handed! 

We hope you are convinced now about the fundamental nature of left and right
at the quark and lepton level. Moreover, the difference between them should 
be completely conventional because we arbitrarily had called left to the one 
$SU(2)$ factor of the Lorentz group and right to the another $SU(2)$ factor.
So one expects the world to be left-right symmetric. Parity interchanges 
left and right. Therefore parity invariance of the world is also expected. 
But we know that the weak interactions are not parity invariant and the
world reflected in a mirror looks different from the original: 
the {\bf P}-mirror image of the left-handed neutrino is right-handed neutrino
which is not observed experimentally.

But this absence of right-handed neutrino not yet indicates left-right 
asymmetry of the world. It is certainly true that under space inversion left
and right becomes interchanged. But in presence of some internal symmetry,
when there are several equivalent left-handed states and several equivalent
right-handed states, it is not obvious at all what right-handed state should
correspond to a given left-handed state. A priori all operators of the type
{\bf PS}, where {\bf P} is the (naive) parity operator considered above and
{\bf S} is some internal symmetry operator, are equally good to represent 
space inversion. Usually internal symmetry {\bf S} is broken. But the parity
symmetry {\bf P} is also broken as we have seen above. So it may happen that
{\bf PS} remains unbroken nevertheless and therefore it can be served as 
representing the equivalence between left and right. What remains is to find
a good enough internal symmetry {\bf S}. And the charge conjugation 
{\bf C}, that is the symmetry between particles and anti-particles, is an
obvious candidate \cite{10}. Indeed the world looks symmetric when 
reflected in the {\bf CP}-mirror because under {\bf CP} left-handed neutrino
goes into right-handed antineutrino and vice versa.     

But {\bf CP} invariance is also broken as experiments in the neutral kaon
system had shown. Recent experiments in the neutral B-meson system also
indicate that our world is not {\bf CP}-symmetric and therefore it is 
either left-right asymmetric or {\bf CP} does not represent the symmetry 
between left and right. Most of the scientific community accepted the first
possibility of the left-right asymmetric world after the remarkable discovery
of {\bf CP} violation in K-meson
decays. This viewpoint remains dominant today. But it is not necessarily
correct. Nineteenth century humorist Josh Billings warned long ago \cite{11}
"The trouble with most folks isn't so much their ignorance. It's know'n so 
many things that ain't so." Evolution of physics is a subtle interplay 
between theory and experiment governed by Lee's two laws \cite{12}.  
The rigidity of accepted opinions in physics is well explained by the first
law which says ``Without experimentalists, theorists tend to drift.'' So one
needs breakthrough experiments to change an orthodox view of the world. The
experimental discovery of the {\bf CP} violation in K-meson decays was one
such breakthrough experiment which changed the previous beliefs by a new 
orthodoxy that only the proper Poincar\'{e} symmetries are symmetries of  
Nature and that the improper Poincar\'{e} symmetries, like space inversion 
and time reversal, are violated. But ``Without theorists, experimentalists
tend to falter'' according to the Lee's second law of physicists.
Lee himself provides \cite{12} a classical example to illustrate this 
second law. During two decades a dozen of experiments were performed to 
measure the Michel parameter $\rho$ in $\mu$-decay. Never the new experimental
value lied outside the error bars of the preceding one. Nevertheless  
conclusions about the nature of weak interactions changed dramatically: the
first experiments indicated $\rho=0$ while the final value was $\rho=3/4$,
and the experiments converged to this final value only after the theoretical
prediction. Of course nobody doubts that {\bf CP} and {\bf P} violations are
firmly established experimentally. But, opposite to the common belief, this 
fact does not necessarily means that Nature is left-right asymmetric. The
theoretical idea which rescues left-right symmetry was put forward by  Lee 
and Yang \cite{13} and involves a dramatic duplication of the world. For any 
ordinary particle the existence of the corresponding ``mirror'' particle is 
postulated. These mirror particles are sterile with respect to the ordinary 
gauge interactions but interact with their own set of mirror gauge particles.
Vice versa, ordinary particles are singlets with respect to the mirror gauge 
group which is an exact copy of the Standard Model
$G_{WS}=SU(3)_C\otimes SU(2)_L \otimes U(1)_Y$ group with only
difference that left and right are interchanged. Hence the mirror weak 
interactions reveal an opposite ${\bf P}$-asymmetry so that ${\bf MP}$, where
the internal symmetry operator ${\bf M}$ interchanges ordinary and mirror 
particles, remains unbroken. The world so extended will look symmetric when 
reflected in the ${\bf MP}$-mirror.

Therefore the desired invariance of the world with regard to the space 
inversion operation (and other improper Poincar\'{e} symmetries), combined 
with the experimental fact that {\bf P} and {\bf CP} are broken in our world,
provides strong motivation for the mirror matter idea. One can even imagine
a reason why gauge and matter contents of our world is duplicated. It may 
happen that the low energy world we are familiar with looks quite different
from the world at high energies. Recently more and more popularity gains the
Brane World idea \cite{14}. According to this idea our low energy world is
located on some 3-dimensional wall (a brane) in higher dimensional space.
More precisely, only the gauge and matter particles are located on the brane.
Gravity, in contrast, propagates into a full bulk and so is essentially high
dimensional. But this high dimensionality of gravity is hidden at distances
large compared to the size of extra dimensions. The localization of particles
on the brane is not absolute: it takes place only at low energies. When 
energies are high enough compared to some characteristic scale the 
localization, as well as the brane, disappears and the world restores its
high-dimensional nature for all degrees of freedom. A good illustration of
the Brane World idea is M.~C.~Escher's lithograph ``Liberation'' shown in
Fig.1.
\begin{figure}[htb]
  \begin{center} \mbox{\epsfig{figure=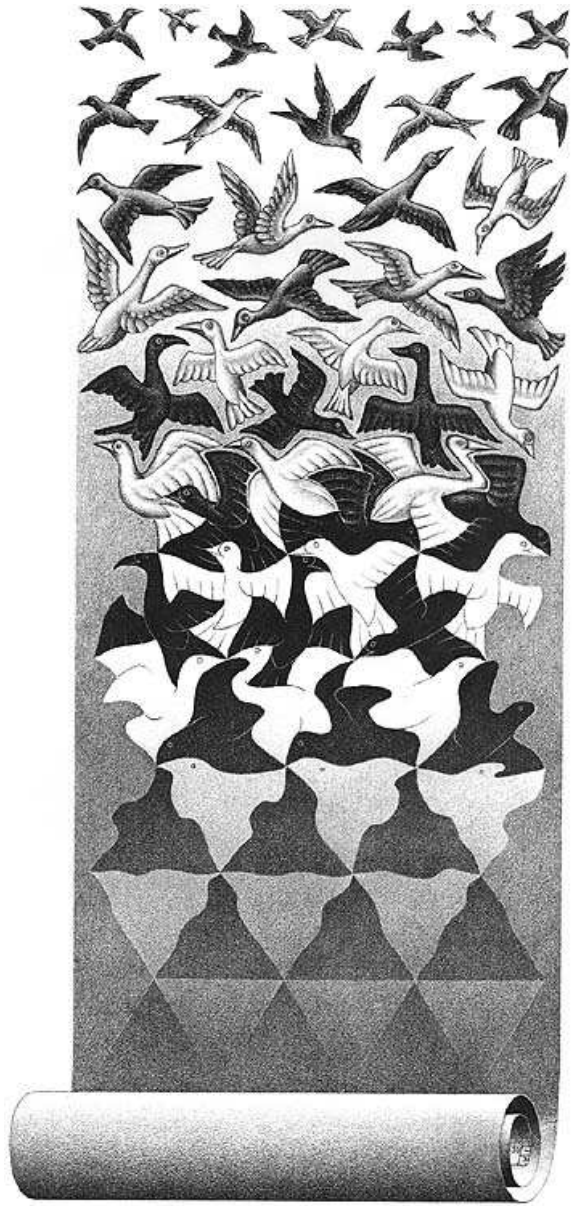 , height=16.0cm}}
  \end{center}
  \caption {The Brane World idea illustrated by M.~C. Escher's lithograph
  ``Liberation''.}
  \label{Fig1}
\end{figure}

Now suppose that the particles can not penetrate (for low energies) the 
thick enough brane and so are localized on one of its surface. Particles 
trapped on the another surface of the brane will appear as some kind of
dark matter for us because objects located on different brane surfaces 
are connected only by gravity, which can penetrate the brane. The parity
invariance of such world could be restored if the parity transformation
involves a transition from one brane surface to another. Therefore the 
mirror particles could be just particles located on the another surface of 
our brane. It is even possible to imagine the mirror world without mirror 
particles if our brane is some analog of  one-sided closed surface. In
\cite{22} this idea was illustrated  
by M.~C.~Escher's woodcut ``M\"{o}bius Strip II''.
Such M\"{o}bius world could be locally left-right asymmetric but 
nevertheless globally
symmetric. For example, suppose a 2-dimensional M\"{o}bius world is 
inhabited by some creatures all having their hearts on left side. So the
symmetry between left and right is violated at least on creatures level.
But inspecting the things closer the creatures will find that this violation
of symmetry is only apparent as there are some shadow mirror creatures 
around which have their hearts on right side. But this mirror creatures
are difficult to discover because they reveal themselves only trough gravity.
So scientists from this 2-dimensional world will need a mirror world to
describe physics around them. From our 3-dimensional perspective, however,
it is quite evident that there is no global difference between ``ordinary''
and ``mirror'' creatures.

We hope your are convinced now that the mirror world idea has good 
theoretical motivation and is not so exotic as it may appear. Maybe some 
historical remarks would be appropriate here. As we had already remarked
the original idea dates back to Lee and Yang's seminal 1956 paper \cite{13}.
But, in contrast to the main conclusion of this paper that our world could
be parity non-invariant, the mirror world way of restoring the symmetry
between left and right had virtually no impact on contemporary science for
a long time. For our best knowledge, for the first time the mirror world
idea was taken seriously and investigated by Kobzarev, Okun and Pomeranchuk
in 1966 paper \cite{15}. Note by the way that the nickname ``mirror world'' 
was firstly coined in this very work. Nevertheless the idea remained unknown
for the majority of researchers as witnessed by the fact that it has been 
rediscovered at least two times \cite{16,17}. But there was a small group
of physicists which knew about the idea and tried to elaborate it. For 
example, the astrophysical consequences of the idea were thoroughly 
investigated by Blinnikov and Khlopov \cite{18}. Recently the idea became
somewhat more popular after in two 1995 papers by Foot and Volkas \cite{19},
Berezhiani and Mohapatra \cite{20} it was noticed that observed neutrino
mysteries (solar neutrino deficit, the atmospheric neutrino problem, Los 
Alamos evidence for neutrino oscillations) can have their roots in mirror 
world. More details and other references on the subject can be found in 
semi-popular reviews \cite{21,22}.

Is there any observational evidence for mirror matter? Certainly there are
no experimental facts which unavoidably demand mirror matter existence. 
Otherwise you would know about this form of matter before our conference.
But there are impressive amount of facts which can be interpreted as 
indicating towards the mirror world. 

The main difficulty in observing mirror
matter is caused by its very weak connection with ordinary matter. These two
forms of matter interact predominantly by gravity only. And gravity is
very feeble interaction. Nevertheless if some mirror object is massive enough
its gravitational effects could be observable. Interestingly one of main 
problems of modern astrophysics is the presence of huge amount of invisible
dark matter in the universe. And the mirror matter is a natural candidate for
dark matter \cite{18,23}. Large clumps of mirror matter will cause 
gravitational lensing effect on the light from background galaxies. And 
recent weak microlensing studies \cite{24} had really discovered two such 
galaxies (or galaxy clusters) almost empty from the luminous matter. Maybe 
this is the first observation of mirror galaxy \cite{21}. Mirror stars in our
galaxy can also produce gravitational microlensing effect on background stars.
Such microlensing effects also had been observed and can be interpreted as 
observations of mirror stars in the Milky Way halo \cite{25}. 

On smaller scales ordinary and mirror matter are expected to be naturally
segregated because they do not have common dissipative interactions 
\cite{18}. So we expect that systems of the solar system size will have
almost definite mirrority. But some small admixture of matter with opposite
mirrority is also not excluded and one expects the existence of binary 
systems like ordinary star with mirror planet or vice versa. Remarkably
some extra-solar planets recently discovered have strange properties like
being very close to their host stars and therefore may be mirror planets
\cite{26,27}. Even more impressive is recent discovery \cite{28} of floating
planets which have no apparent host stars. Instead of being really isolated,
which is unexpected in conventional theories of planet formation, these
``planetary mass objects'' could be ordinary planets orbiting invisible
mirror stars \cite{29}. 

As we see the mirror world model makes at least five predictions about 
gravitational effects which are really observed:
\begin{itemize}
\item the existence of dark matter
\item gravitational lensing effects caused by invisible galaxies
\item microlensing events due to invisible stars
\item strange extrasolar planets
\item isolated planets
\end{itemize}
So this model could be considered as extremely successful. Nevertheless 
there is no conclusive proof that any of the above listed effects is really
some manifestation of mirror world and can not be explained otherwise. 
Further work is needed to establish unambiguously whether the mirror world
really exists. Meanwhile we can speculate about possible mirror solar 
companion(s) \cite{22,32}.

So far we talked about revealing mirror matter through its gravitational 
fingerprints. But gravity is not necessarily the only way to connect the two
worlds. For neutral particles like Higgs, photon and neutrinos one can imagine
ordinary-mirror mixing which is good from the point of modern field theory
(that is the mixing turns out to be gauge invariant and renormalizable). The
mirror world model with these mixing terms predicts three major effects in
particle physics beyond the Standard Model and two of them are really 
observed! The third one involves 
Higgs-mirror Higgs mixing which can modify significantly the Higgs scalar 
properties \cite{17} but we will be able to test this prediction only after
the Higgs discovery.

Neutrino-mirror neutrino mixing leads to maximal neutrino-mirror neutrino
oscillations no matter how small the Lagrangian mixing parameter is. This 
maximality of mixing is a quite general consequence of {\bf MP} symmetry and
provides a clear experimental signature of this model \cite{19}. It seems that
neutrino oscillations (very likely maximal!) are really observed 
experimentally. But unfortunately the mirror world is again slipping away from
our hands: the last experimental data disfavors the pure active-sterile
oscillations \cite{38}. But we do not agree that at present a sterile neutrino
is excluded by experiment and bet that the sterile neutrino will strike
back soon. It seems that the observed neutrino anomalies are more
complex phenomena than it was initially thought. Besides active-sterile 
mixings, neutrinos could have mixings among active (ordinary) species, like
flavor mixing in quark sector, and among their mirror (sterile) partners.
So we do not expect the two flavor active-sterile neutrino oscillations (which 
is excluded now) to be the only loophole for sterile neutrinos. 
But again we have no definite experimental manifestation of the 
mirror world through neutrino oscillations at present -- only indications 
towards it.

Photon-mirror photon mixing if present will result in a small ordinary 
electric charge acquired by mirror charged particles. As a result mirror 
matter will be able to interact with ordinary matter electromagnetically
although by much reduced strength. But compared to gravity the electromagnetic
interactions are tremendously powerful. So even a very small mixing can lead
to interesting observable effects. For example, orthopositronium will mix 
with mirror orthopositronium and decay into an invisible state \cite{39}.
So its decay rate will not coincide with the theoretical prediction \cite{39}.
Interestingly such discrepancy is really observed in some experimental 
measurements and the mirror world may help to resolve this longstanding
discrepancy \cite{40}. Note that the mirror world effect on the 
orthopositronium decays will appear only in vacuum experiments because 
otherwise the ordinary matter environment will destroy coherence between 
mirror and ordinary parts of the orthopositronium state vector and suppress
the oscillations. Such kind of coherence loss is important also in other 
phenomena involving photon-mirror photon mixing \cite{41}.
 
The resolution of the orthopositronium lifetime puzzle via mirror world 
scenario requires relatively large  photon-mirror photon mixing parameter
\cite{40}. If the mixing parameter is indeed so large an interesting 
possibility will be opened that the mirror world can lead to the Tunguska-like
events and maybe the Tunguska event itself was a manifestation of the mirror
world \cite{2}. But this is another story -- for details see Foot's talk at
this conference \cite{2}. Instead we will speculate now about a possibility 
that a tiny electromagnetic interaction between the mirror and ordinary atoms
will be nevertheless enough to prevent ordinary accretion material near the
large mirror body from falling to its center. If the repulsion between 
ordinary and mirror electron orbitals is enough to overcome gravitational 
attraction on the surface of mirror body the ordinary dust and other accreted
material will stay on the surface and will form some kind of very fragile and
porous crust. As a result the mirror body will become visible and may appear
as some strange object for a distant observer. Are there any such objects in
the solar system? We speculated earlier \cite{22,32} that  there might be a 
mirror planet in our solar system that might be found one day. But perhaps an 
even more interesting possibility is that one has already been found! There 
are some strange objects observed in the solar system and we list them below.
Potentially they could be candidates for a mirror celestial body covered by
ordinary crust.

Let us begin with the ninth planet Pluto \cite{Foot}. Some of its strange 
properties are
\cite{42}:   
\begin{itemize}
\item highly eccentric orbit
\item orbital inclination much higher than the other planets'
\item the second most contrasty body in the solar system
\item covered with exotic, super-volatile snows of nitrogen, methane and 
carbon monoxide
\end{itemize} 
There is also some evidence that the Pluto's surface is very porous \cite{43}.
Maybe one day NASA can send space ships to Pluto to bring back mirror matter! 
That could be useful because the mirror matter should have useful industrial 
applications \cite{2}. 

Another strange object in the solar system is Saturn's outermost satellite 
Phoebe. Here are some of its oddities \cite{44}:
\begin{itemize}
\item very low albedo, it is as dark as lampblack
\item eccentric, retrograde orbit
\item high orbital inclination
\item anomalously low density of about $0.7~\mathrm{g/cm^3}$
\end{itemize}
Of course, rather being a mirror object, phoebe more likely may be a dark
carbonaceous captured asteroid formed in the outer solar system as scientists
believe. But who knows \ldots

\begin{figure}[htb] 
   \begin{center} \mbox{\epsfig{figure=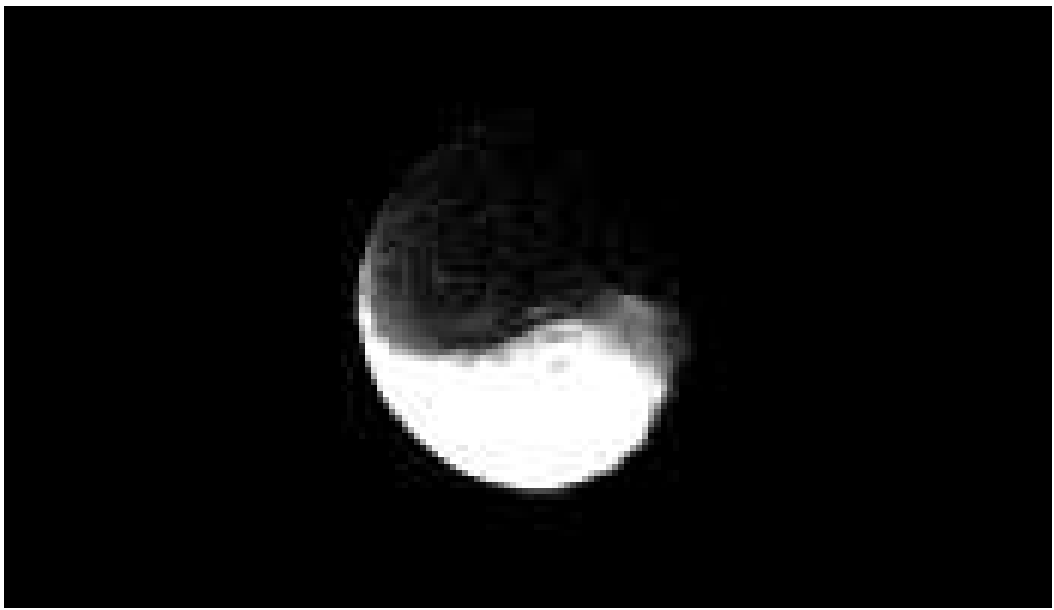 , height=8.0cm}} 
\caption{Iapetus' image taken by Voyager-1 spacecraft}
\label{Fig5}
     \end{center}   
 \end{figure}

Undoubtedly the strangest object in the solar system and maybe the best 
candidate for mirror object covered by ordinary crust is Saturn's another
moon Iapetus \cite{45}. It orbits not in a plane of the other moons. Its 
density $1.1~\mathrm{g/cm^3}$ indicates that Iapetus must be composed almost
entirely of water ice. Indeed its trailing hemisphere is very bright. But 
mysteriously the leading hemisphere is completely different -- it is as dark 
as lampblack. See Fig.\ref{Fig5} which shows Iapetus as seen by Voyager-1 
spacecraft. 
Standard explanation is that the leading hemisphere is coated by a dust 
material knocked out of Phoebe by meteor impacts. But in this case meteor
impacts on Iapetus dark hemisphere is expected to produce craters with bright
floor and none of them is observed. What is observed is just opposite: 
dark-floored craters in Iapetus' high-albedo hemisphere.   
  
So it is even possible that mirror matter has already been discovered and 
everybody can look at mirror body in our solar system by telescope. But we 
feel we are becoming too open-minded here. So it is good time to finish 
before our brains fall out.

\section*{Acknowledgment} 
%\vskip 0.4cm
%\noindent
%{\bf Acknowledgment}
%\vskip 0.4cm
%\noindent
The author thanks A.~Yu.~Ol'khovatov and B.~U.~Rodionov for kind hospitality
during the conference. This paper would be never written without discussions
with Robert Foot. Some fragments from his letters are literally used in the
text. I regret that he refused to be co-author of this paper but understand
that his decision was dictated by high moral standards he adhere to.


\begin{thebibliography}{99}
\bibitem{Foot} The idea that Pluto might be mirror planet belongs to R.~Foot,
as well as a general idea about the possible existence of mirror objects with
ordinary crust.
\end{thebibliography}
\end{document}